\documentstyle[11pt,paspconf,psfig2]{article}

\begin{document}

\title{Imaging Circumstellar Debris Disks}

\author{Ray Jayawardhana\altaffilmark{1}}
\affil{Harvard-Smithsonian Center for Astrophysics, Cambridge, MA 02138, U.S.A.}

% Notice that some of these authors have alternate affiliations, which
% are identified by the \altaffilmark after each name.  The actual alternate
% affiliation information is typeset in footnotes at the bottom of the
% first page, and the text itself is specified in \altaffiltext commands.
% There is a separate \altaffiltext for each alternate affiliation
% indicated above.

\altaffiltext{1}{Visiting Astronomer, Cerro Tololo Inter-American Observatory. 
CTIO is operated by AURA, Inc.\ under cooperative agreement with the National
Science Foundation} 

% The abstract is entered in a LaTeX "environment", designated with paired
% \begin{abstract} -- \end{abstract} commands.  Other environments are
% identified by the name in the curly braces.

% Poster authors ONLY may omit the abstract in order to gain a little
% more page space for the text of the poster.

%\begin{abstract}

%\end{abstract}

% Keywords should be included, but they are not printed in the hardcopy.


% That's it for the front matter.  On to the main body of the paper.
% We'll only put in tutorial remarks at the beginning of each section
% so you can see entire sections together.

\section{Introduction}
In the early 1980s, the Infrared Astronomy Satellite (IRAS) detected
thermal emission from dust grains with temperatures of 50-125 K and 
fractional luminosities ($L_{grains}/L_{*}$) in the range 
$10^{-5}$-$10^{-3}$ around four main sequence stars: Vega, Fomalhaut, 
$\beta$ Pictoris, and $\epsilon$ Eridani. Coronagraphic observations 
of $\beta$ Pic confirmed that the grains do indeed lie in a disk, 
perhaps associated with a young planetary system (Smith \& Terrile 1984). 
Subsequent surveys of IRAS data have revealed
over 100 other main sequence stars of all spectral classes with 
far-infrared excesses indicative of circumstellar disks (Aumann 1985; 
Sadakane \& Nishida 1986; Cote 1987; Jascheck et al. 1991; Oudmaijer et al. 
1992; Cheng et al. 1992; Backman \& Paresce 1993; Mannings \& Barlow 1998). 

In most ``Vega-type'' stars, the dust grains responsible for the infrared 
emission are thought to be continually replenished by 
collisions and sublimation of larger bodies, because the timescales for grain
destruction by Poynting-Robertson (PR) drag and ice sublimation are much
shorter than the stellar main sequence lifetimes (Nakano 1988; Backman 
\& Paresce 1993). In other words, the disks around main sequence stars are 
likely to be debris or remnant disks rather than protoplanetary structures.
These debris disks contain much less dust (and gas) than the massive,
actively accreting disks frequently observed around young pre-main-sequence 
stars (e.g., Strom et al. 1989; Beckwith et al. 1990). Thus, it is likely 
that circumstellar disks evolve from masssive, optically 
thick, actively-accreting structures to low-mass optically thin structures 
with inner holes, and that disk evolution is closely linked to planet 
formation (For a popular review, see Jayawardhana 1998).

\section{Recent Discoveries}
For nearly 15 years, only one debris disk --that around $\beta$ Pic-- had
been imaged. Coronagraphic surveys at optical and near-infrared
wavelengths had failed to detect scattered light from other Vega-type disks
(Smith et al. 1992; Kalas \& Jewitt 1996). However, recent advances in 
infrared and sub-millimeter detectors have led to a batch of dramatic 
new discoveries.

Using the OSCIR mid-infrared camera on the 4-meter telescope at the Cerro
Tololo Inter-American Observatory in Chile, we recently imaged a dust disk 
around the young A star HR 4796A at 18$\mu$m (Jayawardhana et al. 1998; 
Koerner et al. 1998; also see Jura et al. 1998, and references therein). 
HR 4796A is unique among Vega-type stars
in that its low-mass binary companion --at an apparent separation of 500AU--
shows that the system is relatively young. With a variety of constraints such 
as lithium abundance, rotational velocity, H-R diagram position and coronal 
activity, Stauffer et al. (1995) infer that the companion is only $8\pm3$ Myrs 
old, an age comparable to the $\sim$10-Myr timescale estimated for planet
formation (Strom, Edwards, \& Skrutskie 1993; Podosek \& Cassen 1994). 
Interestingly, mid-infrared images of the HR 4796A disk do indeed suggest
the presence of an inner cavity of solar system dimensions, as one would
expect if dust in that region has coagulated into planets or planetesimals.

\begin{figure}
{\psfig{figure=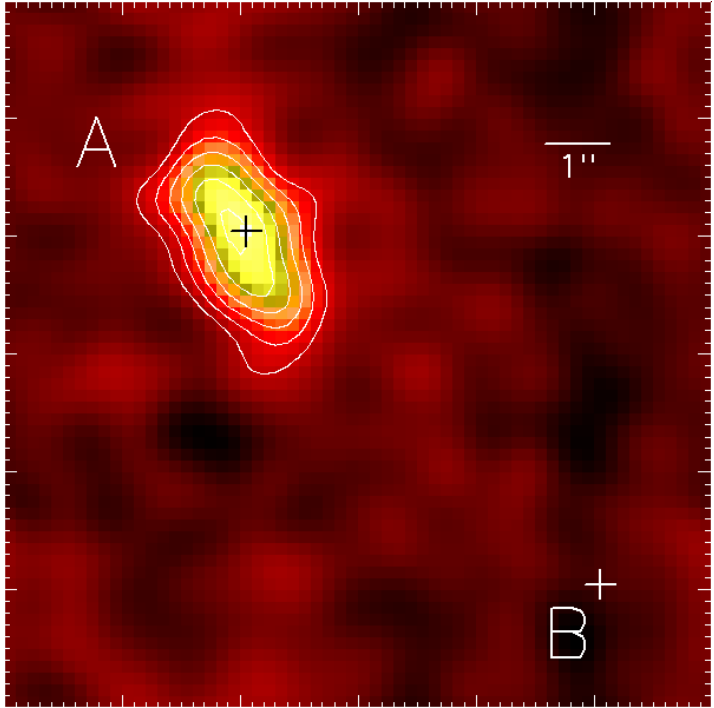,height=3.7in,width=3.7in}}
\caption{The HR 4796A dust disk at 18$\mu$m.} 
\end{figure}

Using a revolutionary new camera known as SCUBA on the James Clerk Maxwell
Telescope in Hawaii, the four prototype debris disks --those around Vega,
$\beta$ Pic, Fomalhaut and $\epsilon$ Eri-- have also been resolved for
the first time in the sub-millimeter (Holland et al. 1998; 
Greaves et al. 1998). These disks, whose parent stars are likely to be
older than HR 4796A, also show inner holes, which may persist due to 
planets within the central void consuming or perturbing grains inbound 
under the influence of the PR drag (Holland et al. 1998; Greaves et al. 
1998). A surprising result is the discovery of giant ``clumps'' 
within three of the disks. The origin and nature of these 
apparent density enhancements remain a mystery.

Another exciting new result is the discovery of dust emission around
55 Cancri, a star with one, or possibly two, known radial-velocity planetary
companions (Butler et al. 1997). From observations made by the Infrared 
Space Observatory
(ISO) at 25$\mu$m and 60$\mu$m, Dominik et al. (1998) concluded that
55 Cancri, a 3-5 billion year old G8 star, has a Vega-like disk
roughly 60 AU in diamater. Recent near-infrared coronographic observations
of Trilling \& Brown (1998) have resolved the scattered light from the 55 
Cancri dust disk and confirm that it extends to at least 40 AU 
(3.24'') from the star. Their findings suggest that a significant amount of
dust --perhaps a few tenths of an Earth mass-- may be present even in
the Kuiper Belts of mature planetary systems.

\section{What next?}
While these remarkable new findings have given us unprecedented glimpses
into circumstellar debris disks at a variety of stages and conditions, 
the timescale for disk evolution is still highly uncertain, and may not even
be universal. The primary obstacle to determining evolutionary timescales is
the difficulty in assigning reliable ages to isolated main sequence stars. 
Fortunately, as in the case of HR 4796, it is possible to obtain
reliable ages for Vega-type stars which have low-mass binary or common 
proper motion companions (Stauffer et al. 1995; Barrado y Navascues et 
al. 1997). Therefore, if we are able to image disks surrounding 
a sample of Vega-type stars whose ages are known from companions, it may 
be possible to place them in an evolutionary sequence, and to constrain 
the timescale(s) for planet formation. 

The recent identification of a group of young stars associated with TW Hydrae
offers a valuable laboratory to study disk evolution and planet formation
(Kastner et al.1997; Webb et al. 1998). Being the nearest group of young 
stars, at a distance of $\sim$55 pc, the TW Hydrae Association is ideally
suited for sensitive disk searches in the mid-infrared. Furthermore, its
estimated age of $\sim$10 Myr would provide a strong constraint on disk
evolution timescales and fill a significant gap in the age sequence between
$\sim$1-Myr-old T Tauri stars in molecular clouds like Taurus-Auriga and
Chamaeleon and the $\sim$30-Myr-old open clusters such as IC 2602 and
IC 2391. Since several of the TW Hya members are close binary
systems, it will also be possible to study the effects of companions on disk
evolution. It could well be that a companion's presence dramatically 
accelerates disk depletion or causes disk assymmetries.

The current mid-infrared cameras when used on Keck should also be able to
conduct sensitive searches for thermal emission from dust associated with 
Kuiper Belts of extrasolar planetary systems like 55 Cancri. In the
sub-millimeter, the beam sizes are still too large to spatially resolve
such disks, but sub-mm flux measurements would place important 
constraints on the amount and nature of their dust.

We are just beginning to study the diversity of debris disks, their 
evolution and their close connection to planets. New instruments like 
the Sub-Millimeter Array (SMA) and upcoming space missions such as the
Space Infrared Telescope Facility (SIRTF) will no doubt assist in that 
quest.

% Finally, we have a little acknowledgements section.

\acknowledgments
I am most grateful to my collaborators Lee Hartmann, Giovanni Fazio,
Charles Telesco, Scott Fisher and Robert Pi\~na. It is also my pleasure 
to acknowledge useful discussions with David Barrado y Navascues, 
Wayne Holland, Jane Greaves, Geoffrey Marcy, and David Trilling.

% That's the end of the main body of the paper.  Now we will have some
% back matter.

% Now comes the reference list.  Since we typed out the citations ourselves,
% the reference list is enclosed in a "references" environment.  Each
% new reference begins with a \reference command which sets up the proper
% indentation.  Typography that may be required in the reference list by
% the editorial staff must be included by the author.
%
% Observe the "standard" order for bibliographic material: author name(s),
% publication year, journal name, volume, and page number for articles.
% Some journal names are available as macros; see the WGAS markup
% instructions for a listing of which ones have been "macro-ized".
% Note the use of curly braces to delimit the font changes: it is essential
% that this be done to limit the scope of the font declaration.
%
% There is no need to engage in any other typographic manipulation.

% That's all, folks.
%
% The technique of segregating major semantic components of the document
% within "environments" is a very good one, but you as an author have to
% come up with a way of making sure each \begin{whatzit} has a corresponding
% \end{whatzit}.  If you miss one, LaTeX will probably complain a great
% deal during the composition of the document.  Occasionally, you get away
% with it right up to the \end{document}, in which case, you will see
% "\begin{whatzit} ended by \end{document}".

\end{document}